\documentclass[12pt]{article}
\usepackage{amssymb,amsmath,epsfig}

\begin{document}

\title{\bf Effects of CDTT Model on the Dynamical Instability of
Cylindrically Symmetric Collapsing Stars}

\author{Hafiza Rizwana Kausar
\thanks{rizwa\_math@yahoo.com}\\\\
Centre for Applicable Mathematics \& Statistics, \\
Business School, University of Central Punjab,\\ Lahore-Pakistan.}
\date{}
\maketitle

\begin{abstract}
We assume cylindrically symmetric stars which begin collapsing by
dissipating energy in the form of heat flux. We wish to study the
effects of Carroll-Duvvuri-Trodden-Turner (CDTT) model,
$f(R)=R+\sigma\frac{\mu^4}{R}$, on the range of dynamical
instability. For this purpose, perturbation scheme is applied to all
the metric functions, material functions and $f(R)$ model to obtain
the full set of dynamical equation which control the evolution of
the physical variables at the surface of a star. It is found that
instability limit involves adiabatic index $\Gamma$ which depends on
the density profile and immense terms of perturbed CDTT model. In
addition, model is constrained by some requirement, e.g. positivity
of physical quantities. We also reduce our results asymptotically as
$\mu\rightarrow0$, being the GR results in both the Newtonian and
post Newtonian regimes.
\end{abstract}

\section{Introduction}

An increasing attention has been paid to the modification of
Einstein-Hilbert (EH) action with the lagrangian density
$\sqrt{-g}R$, where $g$ is the metric tensor and $R$ is the
curvature scalar. A simplest modification is to add terms which are
proportional to $\sqrt{-g}R^n$. It is known that for $n>1$, terms
help to understand the standard cosmology at early times showing
de-Sitter behavior \cite{Starobinsky}. For $n<0$, these corrections
provide a possible gravitational alternative to dark energy by
self-accelerating vacuum solutions and hence lead to the cosmic
speed-up \cite{CDTT}.

The most studied, simplest and definiteness $f(R)$ models in the
literature are CDTT model $f(R)=R+\sigma\frac{\mu^4}{R}$, super
gravity model $f(R)=R+\alpha R^2$ and their mixture generalized CDTT
model $f(R)=R+\sigma\frac{\mu^4}{R}+\alpha R^2$, where $\sigma=\pm1$
$\alpha$ is a positive real number and $\mu$ is a parameter with
units of mass. Both the CDTT and generalized CDTT models allow
de-sitter solution with $R_0=\sqrt{3}\mu^2$, whenever $\mu>0$. By
choosing $\mu\sim H_0$, the present accelerating expansion can be
achieved and even early time cosmological evolution can be
recovered. Since laws of gravity gets modified on large distances in
$f(R)$ models, this allows several interesting observational
signatures such as modification to the spectra of the galaxy
clustering \cite{cb1, cb2}, cosmic microwave background \cite{zs1,
zs2} and weak lensing \cite{st1, st2}.

In order to provide $f(R)$ gravity as a consistent theory of
gravity, $f(R)$ models are severely constrained cosmologically and
by gravitational physics \cite{Mena1}-\cite{Mena3}. The best known
scale of gravitational physics is the solar system scale. To have
stable stellar configuration, the function $f(R)$ should satisfy the
condition, $d^2f/dR^2\geq0$ at each point inside the star
\cite{HuW}. If this condition is not fulfilled, the star would
collapse to some new configuration due to small perturbation. Also,
this theory should compared to parameterized post-Newtonian (PPN)
approximation, where the non-trivial metric elements outside the
stars are $g_{00}=1-\frac{2m}{r}$ and
$g_{11}=\frac{1}{1-\gamma_{PPN}\frac{2m}{r}}$ with observational
value $\mid \gamma_{PPN}-1\mid \ll1$. However, most of the $f(R)$
models gives $\gamma_{PPN}=\frac{1}{2}$, contradicting the
observations.

Gravitation theory and relativistic astrophysics have gone through
extensive developments to the discovery of stellar collapse. A star
has a life cycle wherein it is born in gigantic clouds of dust and
galactic material, then evolve and shine for millions of years and
eventually enter the phase of dissolution and extinction. It happens
when internal pressures subside by nuclear process, gravity takes
over, and the star begins to contract and collapses onto itself. Due
to the achievements of $f(R)$ gravity in numerous areas of cosmology
and astrophysics, it is quite natural to discuss dynamics of
gravitational collapse in this theory. In recent years, we have
explored some aspects of dark energy and gravitational collapse in
$f(R)$ theory \cite{Riz1}-\cite{Riz9}.

The problem of dynamical instability is closely associated with
formation and evolution of self-gravitating objects. The pioneer
work in this direction was done by Chandrasekhar \cite{1}.
Afterwards, this issue has been investigated for adiabatic,
non-adiabatic, anisotropic and shearing viscous fluids by Herrera et
al. \cite{2'}-\cite{4}. In general, dynamical instability discussed
in terms of adiabatic index $\Gamma$. The expression for isotropic
spheres may be given by $\Gamma\geq\frac{4}{3}+n\frac{M}{r}$, where
$n$ is a number of order unity that depends on the structure of a
star, $M$ and $R$ are the mass and radius of a star respectively.
For example, for white dwarfs $n = 2.25$. Instability limit of
collapsing stars also depends upon the physical quantities
describing the properties of different fluids filling the stars. For
example, the dissipating quantities increases the instability range
at Newtonian corrections but makes the fluid less unstable at
relativistic corrections \cite{5}. Also, Chan et al. found
interesting results by studying the effects of anisotropy, radiation
and shearing viscosity at Newtonian (N) and post-Newtonian (PN)
regimes \cite{6}-\cite{8}. Sharif and Azam \cite{19, 20}, have
discussed the effects of electromagnetic field on the dynamical
instability of spherical and cylindrical symmetric gravitational
collapse. We have investigated the effects of a well known
$f(R)=R+\alpha R^2$ model on the dynamical instability of
expansion-free spherically symmetric gravitational collapse
\cite{Riz10}.

It is a fact that most of the collapse models analyzed so far are
spherical due to their wide astrophysical significance. However, the
study of non-spherical collapse remains a major uncharted territory,
even many attempts have been made, e.g., \cite{70, 71}.
Specifically, final fate of collapse of a non-spherical cloud coming
from numerical relativity and certain analytical solutions in
cylindrical symmetry provide some new examples about gravitational
collapse \cite{72, 73}. Nevertheless, a comprehensive analytical
treatment of collapse still needs to be done. In this connection, we
would like to extend our work on dynamical instability to the
cylindrically symmetric collapsing stars in $f(R)$ gravity.

The manuscript is laid out as follows. In Section \textbf{2}, we
provide a full description of the matter distribution, the line
element, both inside and outside the fluid boundary, and the field
equations. We also formulate the dynamical equations which governs
the dynamics of gravitational collapse. In section \textbf{3}, we
consider CDTT model and discuss its feature on cosmology and
gravitation. Also, we apply perturbation scheme to all the function
and $f(R)$ model. Considering Newtonian and post-Newtonian limits
instability range in the form of $\Gamma$ factor is obtained. In the
last section \textbf{4}, work is summarized followed by an Appendix.

\section{Dynamical Equations}

The modified form of EH action leading to $f(R)$ gravity can be
written as
\begin{equation}\label{b}
S=\frac{1}{2\kappa}\int d^{4}x\sqrt{-g}f(R),
\end{equation}
where $\kappa$ is the coupling constant and $f(R)$ is any arbitrary
function of Ricci scalar. The field equations in metric approach are
reached by variating above action with respect to $g_{\alpha\beta}$
as follows
\begin{equation}\label{b'}
F(R)R_{\alpha\beta}-\frac{1}{2}f(R)g_{\alpha\beta}-\nabla_{\alpha}
\nabla_{\beta}F(R)+ g_{\alpha\beta} \Box F(R)=\kappa
T^m_{\alpha\beta},\quad(\alpha,\beta=0,1,2,3),
\end{equation}
where $F(R)\equiv df(R)/dR$, $\Box=\nabla^{\mu}\nabla_{\mu}$ with
$\nabla_{\mu}$ representing the covariant derivative and
$T^m_{\alpha\beta}$ is the standard minimally coupled stress-energy
tensor. Alternatively, this can be written in a form similar to the
field equations of General Relativity (GR)
\begin{equation}\label{12}
G_{\alpha\beta}=\frac{\kappa}{F}[T_{\alpha\beta}^{m}+T_{\alpha\beta}^c],
\end{equation}
where
\begin{equation}\label{d}
T_{\alpha\beta}^c=\frac{1}{\kappa}\left[\frac{f(R)-RF(R)}{2}g_{\alpha\beta}+\nabla_{\alpha}
\nabla_{\beta}F(R) -g_{\alpha\beta} \Box F(R)\right].
\end{equation}
This way of writing field equations help to study the dark side of
the universe in term of curvature terms provided that it does not
satisfy the usual energy conditions. Consequently, this theory may
be used to explain the expansion of the universe and other aspects
of dark energy in the gravitational physics.

We consider a cylindrically symmetric collapsing star bounded by a
hypersurface $\Sigma$. The interior region inside the boundary can
be represented by the following line element
\begin{equation}\label{1}
ds^2_-=A^2(t,r)dt^{2}-B^2(t,r)dr^{2}-C^2(t,r)d\theta^{2}-dz^{2}.
\end{equation}
In order to preserve cylindrical symmetry, following constraints may
be imposed on the coordinates
\begin{equation}\label{1'}
-\infty \leq t \leq \infty,~~0\leq r < \infty,~~-\infty < z <
\infty,~~0 \leq \theta \leq 2\pi.
\end{equation}
The exterior region across the boundary of that star can be
represented by a cylindrically symmetric manifold in the retarded
time coordinate $\nu$ as follows \cite{29}
\begin{equation}\label{25}
ds^2_+=-\frac{2M(\nu)}{r}d\nu^2+2drd\nu-r^2(d\theta^2+\gamma^2dz^2),
\end{equation}
where $\gamma^2=-\frac{\Lambda}{3}$, $\Lambda$ being the
cosmological constant while $M(\nu)$ is the total mass inside the
boundary surface. As gravitational collapse is a highly dissipative
process, we assume that fluid filling the collapsing cylinder is
dissipating energy in the form of heat flux $q$ and can be
represented by the following energy-momentum tensor \cite{5, 20}
\begin{equation}\label{2}
T_{\alpha\beta}=(\rho+p)u_{\alpha}u_{\beta}-pg_{\alpha\beta}+
q_\alpha u_\beta+q_{\beta}u_\alpha,
\end{equation}
where $\rho$ stands for energy density, $p$ for pressure and
$u_{\alpha}$ for four-velocity of the fluid. In co-moving
coordinates, these quantities satisfy the relations
\begin{equation}\label{3}
u^{\alpha}=A^{-1}\delta^{\alpha}_{0},\quad
u^{\alpha}u_{\alpha}=1,\quad q^{\alpha}=qB^{-1}\delta^\alpha_1,\quad
u^{\alpha}q_{\alpha}=0.
\end{equation}
The effects of dissipation describe a wide range of situations. For
example, using quasi-static approximation, limiting cases of
radiative transport have been studied in \cite{52}. It is found that
hydrostatic time scale is very small as compared to the stellar
lifetimes for different phases of a star's life. It is of the order
of 27 minutes for the sun, 4.5 seconds for a white dwarf and
$10^{-4}$ seconds for a neutron star of one solar mass and 10 km
radius \cite{53}.

For interior metric, Eqs.(\ref{12}) yield the following set of field
equations
\begin{eqnarray}\nonumber
&&\left(\frac{A}{B}\right)^2 \left(\frac{B'C'}{BC}-\frac{C''}{C}
\right)=\frac{\kappa}{F}\left[{\rho}A^{2}+\frac{A^2}{\kappa}\left\{\frac{f-R
F}{2}+\frac{F''}{B^2}+\frac{F'}{B^2}\frac{C'}{C}\right.\right.\\\label{13}
&&\left.\left.-\frac{\dot{F}}{A^2}\left(\frac{\dot{A}}{A}
+\frac{\dot{B}}{B}+\frac{\dot{C}}{C}\right)
\right\}\right],\\\label{14}
&&\frac{\dot{C'}}{C}+\frac{A'\dot{C}}{AC}+\frac{\dot{B}C'}{BC}
=\frac{\kappa}{F}\left[qAB+\frac{1}{\kappa}\left(\dot{F}'
-\frac{A'}{A}\dot{F}-\frac{\dot{B}}{B}F'\right)\right],
\end{eqnarray}
\begin{eqnarray}\nonumber
&&\left(\frac{B}{A}\right)^2\left(\frac{\dot{A}}{A}
\frac{\dot{C}}{C}-\frac{\ddot{C}}{C}\right)
+\frac{A'}{A}\frac{C'}{C} =\frac{\kappa}{F}
\left[pB^{2}+\frac{B^2}{\kappa}\left\{-\frac{f-R
F}{2}+\frac{\ddot{F}}{A^2}\right.\right.\\\label{15} &&\left.\left.+
\frac{\dot{F}}{A^2}\frac{\dot{C}}{C}+\frac{F'}{B^2}\left(\frac{A'}{A}+
\frac{B'}{B}+\frac{C'}{C}\right)\right\}\right],\\\nonumber
&&\frac{1}{AB}\left(\frac{A''}{B}-\frac{\ddot{B}}{A}
+\frac{\dot{A}\dot{B}}{A^2}-\frac{A'B'}{B^2}\right)
=\frac{\kappa}{F}\left[p-\frac{1}{\kappa}\left\{-\frac{f-R
F}{2}+\frac{\ddot{F}}{A^2}\right.\right.\\\label{16}
&&\left.\left.-\frac{F''}{B^2}+
\frac{\dot{F}}{A^2}\frac{\dot{B}}{B}-
\frac{F'}{B^2}\frac{A'}{A}\right\}\right], \\\nonumber
&&\frac{A''}{AB^2}-\frac{\ddot{B}}{A^2B}+\frac{\dot{A}\dot{B}}{A^3B}
-\frac{A'B'}{AB^3}+\frac{\dot{A}\dot{C}}{A^3C}
-\frac{\ddot{C}}{A^2C}-\frac{B'C'}{B^3C}+\frac{C''}{B^2C}\\\nonumber
&&+\frac{A'C'}{AB^2C}-\frac{\dot{B}\dot{C}}{A^2BC}
=\frac{\kappa}{F}\left[p+\frac{1}{\kappa}\left\{-\frac{f-R
F}{2}+\frac{\ddot{F}}{A^2}-\frac{F''}{B^2}\right.\right.\\\label{17}
&&\left.\left.+\frac{\dot{F}}{A^2}\left(\frac{\dot{B}}{B}+\frac{\dot{C}}{C}\right)
-\frac{F'}{B^2}\left(\frac{A'}{A}+\frac{C'}{C}\right)\right\}\right].
\end{eqnarray}
Here dot and prime denote derivatives with respect to $t$ and $r$
respectively. Subtracting Eq.(\ref{16}) from (\ref{15}), we have
\begin{eqnarray}\nonumber
&&\frac{1}{A^2}\left(\frac{\dot{A}}{A}-\frac{\dot{B}}{B}
+\frac{\dot{F}}{F}\right)\frac{\dot{C}}{C}
+\frac{1}{B^2}\left(\frac{A'}{A}-\frac{B'}{B}
-\frac{F'}{F}\right)\frac{C'}{C}+\frac{1}{C}
\left(\frac{C''}{B^2}-\frac{\ddot{C}}{A^2}\right)=0.\\\label{18}
\end{eqnarray}

The dynamical equations help to investigate the evolution of
gravitational collapse with time and yield the variation of total
energy inside a collapsing body with respect to time and adjacent
surfaces. We have formulated these equations by using contracted
Bianchi identities both for the usual matter and effective
energy-momentum tensor carrying curvature terms. These are given by
\begin{eqnarray}\label{bb}
&&\left(\overset{(m)}{T^{\alpha\beta}}+\overset{(c)}
{T^{\alpha\beta}}\right)_{;\beta}u_{\alpha}=0,\quad
\left(\overset{(m)}{T^{\alpha\beta}}+\overset{(c)}{T^{\alpha\beta}}\right)_{;\beta}
\chi_{\alpha}=0.
\end{eqnarray}
The two independent components of Bianchi identities for the system
under consideration read
\begin{eqnarray}\nonumber &&\dot{\rho}-q'\frac{A}{B}+q\frac{A}{B}\left(\frac{2A'}{A}
-\frac{C'}{C}\right)+(\rho+p)
\left(\frac{\dot{B}}{B}+\frac{\dot{C}}{C}\right)+\phi_1(r,t)=0,\\\label{19}\\\label{21}
&&p'+\dot{q}\frac{B}{A}+q\frac{B}{A}\left(\frac{2\dot{B}}{B}
+\frac{\dot{C}}{C}\right)+(\rho+p)\frac{A'}{A}+\phi_2(r,t)=0,
\end{eqnarray}
where $\phi_1(r,t)$ and $\phi_2(r,t)$ are given in Eqs.(\ref{19*})
and (\ref{21*}) respectively in \textbf{Appendix}.

\section{CDTT Model and Perturbation Scheme}

In this section, we consider the following form of CDTT model,
\begin{equation}\setcounter{equation}{1}\label{gfr}
f(R)=R+\sigma\frac{\mu^4}{R},
\end{equation}
where $\sigma$ was taken equal to $-1$ in the original paper
\cite{CDTT}. It is claimed that with the parameter $\mu$ chosen as
the order of the inverse of the universe age, $\mu^{-1}\sim 10^{18}$
sec $\sim (10^{33}eV)^{-1}$, this model can be applied to cosmology
and describes the present accelerating expansion of the universe.
However, this additional term was not essential to describe earlier
time inflation.  Further, CDTT model has a homogeneous solution of
de-Sitter metric with constant scalar curvature
$R=R_0\equiv\sqrt{3}\mu^2$. Thus, in order to have the desired
physically plausible late time behavior, we have to set
$\sqrt{3}\mu^2\sim H^2_ 0$.

Henttunen, et al. \cite{Henttunen} have analyzed the properties of
polytropic stars by considering CDTT model with the conclusion that
the density profiles in general resemble the Newtonian Lane-Emden
solutions. They analyzed that interior solution for metric
components and curvature of such stars are always singular,
consequently, $f(R)= R-\mu^4/R$ model is not experimentally suitable
to describe the space-time around the sun. A possible way to avoid
is to relax the requirement data set for the central boundary
conditions, but a more plausible approach is to modify the
functional form of $f(R)$. As it is mentioned in the introduction
that sign of $f''(R)$ determine whether this modified theory
approaches the GR limit or not at high curvature. For $f''(R)>0$,
the model behave very close to GR and is stable and vice versa. Due
to this reason, the original CDTT model was severely criticized
\cite{B, Aw}.

In order to avoid instability, Hu and swaicky introduced CDTT model
with $\sigma=1$ \cite{HuI}. Thus, throughout the paper, we shall use
following form
\begin{equation}\label{fr}
f(R)=R+\frac{\mu^4}{R}.
\end{equation}
For the present model, the stability condition is satisfied with
extremely small and positive $f''(R)$ everywhere inside and near the
polytrope. Also, it aims to explain the current accelerated
expansion of the universe. At least for this model, Einstein
evolution can be recovered for most of the cosmic history,
especially standard epoch of matter domination can be obtained,
providing a sufficiently long time to satisfy observations. It is
mentioned here that the additional inverse term will not
significantly alter standard evolution until today and that the
solution lies well within present constraints from Big-Bang
Nucleosynthesis \cite{Evans}.

It is noticeable that CDTT model adds a perturbative function in the
form of inverse power to the EH action. Thus for such a
modification, the Einstein cosmology may be recovered in the limit
that the perturbation term disappear. This system has great
similarity with perturbation theory \cite{55}. First, we analyze our
work on dynamical instability when $\mu\neq0$, then we reduce our
results asymptotically, $\mu\rightarrow0$, being the Einstein
solution.

In this paper, the perturbation scheme introduced is used to analyze
the instability conditions of the dynamical equations. We assume
that initially all the function whether material or metric have only
radial dependence, i.e., lie in the static equilibrium. However,
afterwards, all these quantities have time dependence in their
perturbation. Taking $0<\epsilon\ll1$, quantities may be written in
the following manner
\begin{eqnarray}\label{41} A(t,r)&=&A_0(r)+\epsilon
T(t)a(r),\\\label{42} B(t,r)&=&B_0(r)+\epsilon T(t)b(r),\\\label{43}
C(t,r)&=&rB(t,r)[1+\epsilon T(t)\bar{c}(r)],\\\label{44}
\rho(t,r)&=&\rho_0(r)+\epsilon {\bar{\rho}}(t,r),\\\label{45}
p(t,r)&=&p_{0}(r)+\epsilon {\bar{p}}(t,r),\\\label{46}
m(t,r)&=&m_0(r)+\epsilon {\bar{m}}(t,r),\\\label{48}
q(t,r)&=&\epsilon {\bar{q}}(t,r), \\\label{49'}
R(t,r)&=&R_0(r)+\epsilon T(t)e(r),\\\label{50'}
f(R)&=&[R_0(r)+2\mu^4R_0^{-1}(r)]+\epsilon
T(t)e(r)[1-2\mu^4R_0^{-2}(r)],\\\label{51'}
F(R)&=&1-\mu^4R_0^{-2}(r)+2\epsilon\mu^4R_0^{-4}(r) T(t)e(r).
\end{eqnarray}

The static configuration of the field equations
(\ref{13})-(\ref{15}) and (\ref{18}) is obtained as
\begin{eqnarray}\nonumber
&&\frac{\kappa}{1-\mu^4R_0^{-2}}[\rho_0+\frac{\mu^4R_0^{-1}}{\kappa}\{1+2R_0^{-2}
(-3R_0^{-1}R_0'^{-2}+R_0'')\}]\\\label{50}
&&=\frac{1}{B_0^2}\left[\left(\frac{B_0'}{B_0}\right)^2-\frac{B_0'}{rB_0}
-\frac{B_0''}{B_0}\right],\\\nonumber
&&\frac{\kappa}{1-\mu^4R_0^{-2}}\left[p_{r0}-\frac{\mu^4R_0^{-1}}{\kappa}
\left\{1-\frac{2R_0^{-2}R_0'}{B_0^2}
\left(\frac{A_0'}{A_0}+\frac{2B_0'}{B_0}
+\frac{1}{r}\right)\right\}\right]\\\label{51}
&&=\frac{1}{B_0^2}\frac{A_0'}{A_0}\left(\frac{B_0'}{B_0}
+\frac{1}{r}\right),\\\nonumber
&&\left(\frac{A_0'}{A_0}-\frac{B_0'}{B_0}
-\frac{2\mu^4R_0^{-3}R_0'}{1-\mu^4R_0^{-2}}\right)
\left(\frac{B_0'}{B_0}+\frac{1}{r}\right)+\frac{2B_0'}{rB_0}
+\frac{B_0''}{B_0}=0.\\\label{52}
\end{eqnarray}
Applying static part of the perturbed system, the first dynamical
equation (\ref{19}) is identically satisfied while (\ref{21}) yields
\begin{eqnarray}\label{59}
p_r'+(\rho_0+p_0)\frac{A_0'}{A_0}+\kappa \phi_{2s}=0,
\end{eqnarray}
where $\phi_{2s}$ is the static part of equation $\phi_2$. The
perturbed configurations of Eq.(\ref{19}) leads to
\begin{eqnarray}\nonumber
\dot{\bar{\rho}}-\bar{q}'\frac{A_0}{B_0}+\bar{q}\frac{A_0}{B_0}
\left(\frac{2A_0'}{A_0}-\frac{B_0'}{B_0}
-\frac{1}{r}\right)&+&\\\label{55}\left[(\rho_0+p_{0})\left(\frac{2b}{B_0}
+\frac{\bar{c}}{B_0}\right)+\kappa \phi_{1p}\right]\dot{T}&=&0.
\end{eqnarray}
Here $\phi_{1p}$ denote the perturbed form of equation $\phi_1$ and
is given in \textbf{Appendix}. Applying perturbation on
Eq.(\ref{14}) and eliminating $\bar{q}$, we have
\begin{eqnarray}\nonumber
\bar{q}&=&-\frac{1-\mu^4R_0^{-2}}{\kappa
A_0B_0}\left[\frac{1}{B_0}\left\{\frac{b+\bar{c}}{r}+b'+\bar{c}'
+\frac{A_0'}{A_0}(b+\bar{c})\right\}\right.\\\label{62}
&+&\left.\frac{R_0^{-4}}{1-\mu^4R_0^{-2}}\left\{2\mu^4(e'+4eR_0^{-1}R_0'
-e\frac{A_0'}{A_0})\right\} +\frac{b}{B_0}\right]\dot{T}.
\end{eqnarray}
Differentiating above equation and substituting in the second and
third term of Eq.(\ref{55}) and then integrating resulting equation
with respect to ``t" , we have
\begin{eqnarray}\label{67}
\bar{\rho}&=&\left[-(\rho_0+p_{r0})\left(\frac{2b}{B_0}+\frac{\bar{c}}{B_0}\right)
+\phi_3(r)\right]T,
\end{eqnarray}
where $\phi_3(r)$ is given in \textbf{Appendix}. Considering the
second law of thermodynamics, we can express a relationship between
$\bar{\rho}$ and $\bar{p_r}$ as the ratio of specific heat by
assuming an equation of state of Harrison-Wheeler type as follows
\cite{ch1,HW}
\begin{equation}\label{69}
\bar{p}=\Gamma\frac{p_{r0}}{\rho_0+p_{r0}}\bar{\rho}.
\end{equation}
Here $\Gamma$ measures the variation of pressure for a given
variation of density. We take it constant throughout the region that
we want to study. Substituting value of $\bar{\rho}$ from
Eq.(\ref{67}) in (\ref{69}), we have
\begin{equation}\label{70}
\bar{p}=-\Gamma p_0\left(\frac{2b}{B_0}+\frac{\bar{c}}{B_0}\right)T
+\Gamma\frac{p_{0}}{\rho_0+p_{0}}\phi_3T.
\end{equation}
The perturbed configurations of second Bianchi identity is
\begin{eqnarray}\label{71}
\bar{p}'+\dot{\bar{q}}\frac{B_0}{A_0}+(\bar{\rho}+\bar{p})
\frac{A_0'}{A_0}+(\rho_0+p_{0})
\left(\frac{a}{A_0}\right)'T+\frac{T}{\kappa}\phi_{2p} =0,
\end{eqnarray}
where $\phi_{2p}$ represents the perturbed part of $\phi_2$ and is
provided in the \textbf{Appendix}. Substituting values of $\bar{q}$,
$\bar{\rho}$ and $\bar{p}$ from Eqs.(\ref{62}), (\ref{67}) and
(\ref{70}) respectively in the above equation, we have
\begin{eqnarray}\nonumber
&&\Gamma\left[-p_0\left(\frac{2b}{B_0}+\frac{\bar{c}}{B_0}\right)
+\frac{p_{0}}{\rho_0+p_{0}}\phi_3\right]_{,1}T+(\rho_0+p_{0})
\left(\frac{a}{A_0}\right)'T\\\nonumber
&-&\frac{A_0'}{A_0}\left[(\rho_0+p_{0}+\Gamma
p_0)\left(\frac{2b}{B_0}+\frac{\bar{c}}{B_0}\right)
-\left(\Gamma\frac{p_{0}}{\rho_0+p_{0}}+1\right)\phi_3\right]T\\\nonumber
&+&\frac{T}{\kappa}\phi_{2p}-\frac{1-\mu^4R_0^{-2}}{\kappa
A_0^2}\left[\frac{1}{B_0}\left\{\frac{b+\bar{c}}{r}+b'+\bar{c}'
+\frac{A_0'}{A_0}(b+\bar{c})\right\}\right.\\\label{72}
&+&\left.\frac{R_0^{-4}}{1-\mu^4R_0^{-2}}\left\{2\mu^4(e'+4eR_0^{-1}R_0'
-e\frac{A_0'}{A_0})\right\} +\frac{b}{rB_0}\right]\ddot{T}=0.
\end{eqnarray}
It is mentioned here that above equation is the required evolution
equation for further analysis. The Ricci scalar curvature is given
by
\begin{eqnarray}\nonumber
R&=&\frac{-2}{B^2}\left(-\frac{A''}{A}+\frac{A'B'}{AB}+\frac{A'C'}{AC}
+\frac{B'C'}{BC}\right)\\\label{R}
&-&\frac{2}{A^2}\left(\frac{\ddot{B}}{B}+\frac{\dot{A}\dot{B}}{AB}
+\frac{\ddot{C}}{C}-\frac{\dot{A}\dot{C}}{AC}+\frac{\dot{B}\dot{C}}{BC}\right).
\end{eqnarray}
The static part of the above equation is obtained as follows
\begin{equation}\label{Rs}
R_0(r)=\frac{-2}{B_0^2}\left[-\frac{A_0''}{A_0}-\frac{B_0''}{B_0}+\frac{2A_0'B_0'}{A_0B_0}
+\frac{A_0'}{rA_0}-\frac{B_0'}{rB_0}+\frac{B_0'^2}{B_0^2}\right].
\end{equation}
The perturbed configuration of the Ricci scalar curvature with the
use of Eq.(\ref{Rs}) yields
\begin{eqnarray}\nonumber
&-&\frac{1}{A_0B_0}\left(a''-\frac{aA_0''}{A_0}-\frac{2bA_0''}{B_0}\right)
-\frac{2}{rB_0^3}\left(b'+\bar{c}'B_0'-\frac{\bar{c}B_0'}{r}-
\frac{3bB_0'}{B_0}\right)\\\nonumber
&+&\frac{e}{2}+\frac{\bar{c}}{r^3}-\frac{1}{A_0B_0^3}\left(a'B_0'+bA_0'-\frac{aA_0'B_0'}{A_0}
-\frac{3bA_0'B_0'}{B_0}\right)\\\label{Rp}
&+&\frac{2\bar{c}''}{rB_0^2}+\frac{1}{B_0^2r^2}\left(\bar{c}'-\frac{b}{B_0}
-\frac{\bar{c}}{r}\right)-\frac{\ddot{T}}{T}\left(\frac{b}{rB_0}
-\frac{2\bar{c}}{r}\right)\frac{1}{A_0^2}=0.
\end{eqnarray}
This equation can also be written as
\begin{equation}\label{66}
\ddot{T}(t)-\phi_4(r) T(t)=0,
\end{equation}
where $\phi_4(r)$ is given in the \textbf{Appendix}. In order to
find the instability range, we assume that all the terms of equation
$\phi_4$ are such that it remains positive. Consequently, the
solution of Eq.(\ref{66}) is obtained as
\begin{equation}\label{68}
T(t)=-e^{\sqrt{\phi_4}t}.
\end{equation}
In the following subsections, we use above value in the dynamical
equation Eq.(\ref{72}) to analyze the evolution of the physical
variables at the surface of the star.

\subsection*{Newtonian Limit}

In this regime, we assume that $\rho_0\gg p_0$ and $A_0=1,~B_0=1$ in
order to fulfill the Newtonian limit. Thus substituting
$A_0'=0,~B_0'=0$ in Eq.(\ref{72}), we obtain
\begin{eqnarray}\nonumber
&-&\frac{1-\mu^4R_0^{-2}}{\kappa}\left[\frac{2b+\bar{c}}{r}+b'+\bar{c}'
-\frac{2\mu^4R_0^{-4}(e'-4eR_0^{-1}R_0')}{1-\mu^4R_0^{-2}}\right]\ddot{T}\\\label{73}
&+&\left[\Gamma\left\{-p_0(2b+\bar{c})\right\}_{,1}
+a'\rho_0+\frac{\phi_{2p(N)}}{\kappa}\right]T=0.
\end{eqnarray}
It is mentioned here that $\phi_{2p(N)}$ represent terms belonging
to the Newtonian regime of perturbed second Bianchi identity.
Substituting value of $T$ from Eq.(\ref{68}), we have
\begin{eqnarray}\label{74}
\Gamma<\frac{a'\rho_0+\phi_5(r)}{[p_0(2b+\bar{c})]'},
\end{eqnarray}
where $\phi_5(r)$ is given in \textbf{Appendix}. It is noticeable
here that adiabatic index depends upon the energy density, pressure
and curvature terms in the Newtonian regime. Thus the collapsing
system would remains unstable until Eq.(\ref{74}) satisfied.

\subsubsection*{Asymptotic Behavior}

When $\mu\rightarrow0$, the expression for $\Gamma$ becomes
\begin{eqnarray}\label{74as}
\Gamma<\frac{a'\rho_0+\frac{\phi_4}{\kappa}\left(\frac{2b+\bar{c}}{r}+b'+\bar{c}'
\right)}{[p_0(2b+\bar{c})]'}.
\end{eqnarray}
This results represents the GR solution.

\subsection*{Post Newtonian Limit}

Here, we include relativistic effects upto order $\frac{m_0}{r}$. In
this case, we have
\begin{eqnarray}\label{pn1}
&&A_0=1-\frac{m_0}{r},~~~~B_0=1+\frac{m_0}{r}, \\\label{pn2}
\Rightarrow~~~&&\frac{A_0'}{A_0}=\frac{m_0}{r(r-m_0)},~~~~\frac{B_0'}{B_0}=\frac{-m_0}{r(r+m_0)}.
\end{eqnarray}
Substituting above values in Eq.(\ref{72}), we obtain
\begin{eqnarray}\nonumber
&&\Gamma>\frac{\frac{m_0}{r^2}\left[(\rho_0+p_0)(2b+\bar{c})-\phi_{3(PN)}\right]
+(\rho_0+p_0)\left[a(1+\frac{m_0}{r})\right]'+\phi_{6}(r)}
{[p_0(2b+\bar{c})(1-\frac{m_0}{r})-\frac{\rho_0\phi_{3(PN)}}{\rho_0+p_0}]'
+[p_0(2b+\bar{c})(1-\frac{m_0}{r})-\frac{\rho_0\phi_{3(PN)}}{\rho_0+p_0}]}.\\\label{PN}
\end{eqnarray}
In the above expression $\phi_{3(PN)}$ denote those terms of
$\phi_{3}$ which belong to PN regime whereas $\phi_6(r)$ is given in
\textbf{Appendix}. Thus, system would be unstable in
PN-approximation as long as the above inequality is satisfied. It
can be observed that how relativistic and curvature terms are
affecting the instability range of collapsing star. Further, in
order to fulfill the dynamical instability condition, we need to
keep all the terms positive. Hence, we assume that all the
quantities involved in the above expression are positive. In
addition, following constraints should be satisfied
\begin{eqnarray}\label{c1}
&&(\rho_0+p_0)(2b+\bar{c})>\phi_{3(PN)},\\\label{c2}
&&p_0(2b+\bar{c})(1-\frac{m_0}{r})>\frac{\rho_0\phi_{3(PN)}}{\rho_0+p_0},
\end{eqnarray}
along with $\frac{m_0}{r}<1$.

\subsubsection*{Asymptotic Behavior}

As $\mu\rightarrow0$, adiabatic index $\Gamma$ has the same
representation. However, in this case $\phi_3$ and $\phi_6$
respectively reduced to
\begin{eqnarray}\nonumber
\phi_3(r)&=&\frac{1}{\kappa}\left[\left(\frac{2b+\bar{c}}{r}
+b'+\bar{c}'+\frac{bm_0}{r^2}\right)_{,1}-\frac{1}{r}\left(\frac{2b+\bar{c}}{r}
+b'+\bar{c}'+\frac{bm_0}{r^2}\right)\right]\\\label{60''as}
&+&\frac{m_0}{r\kappa}\left[\left(\frac{2b+\bar{c}}{r}
+b'+\bar{c}'\right)_{,1}-\frac{2}{r}\left(\frac{2b+\bar{c}}{r}
+b'+\bar{c}'\right)\right], \\\label{61''as}
\phi_6(r)&=&\frac{\phi_4}{\kappa}\left[\frac{3b+\bar{c}}{r}+\frac{3m_0}{r^{2}}+(b'+\bar{c}')
\left(1+\frac{2m_0}{r}\right)\right].
\end{eqnarray}
It is mentioned here that in both the N and PN regimes heat flux is
not affecting limit of dynamical instability.

\section{Summary}

In this paper, we have assumed dissipative fluid configuration of a
non-static cylindrically symmetric collapsing star and discussed the
dynamical instability for the final stages of the star by a
perturbative approach. As study of gravitational collapse could be
used as a paradigm to understand stellar formation. We introduced
this issue in modified $f(R)$ theory of gravity. We have considered
CDTT model which leads standard cosmic history. As short-timescale
instabilities are avoided with extremely small and positive
$f''(R)$, so we have considered modified CDTT model with positive
term $\frac{\mu^4}{R}$. It is worthwhile to mention here that this
model satisfying the condition $f''(R)>0$ have stable high curvature
limits and a well behaved cosmological solution with proper era of
matter domination.

To see the effects of CDTT model on the dynamical instability of
fluid evolution during the collapsing process, we have applied a
perturbation scheme on the field equation and dynamical equations.
It is worth mentioned here that recently we have worked on
spherically symmetric gravitational collapse evolving under the
expansionfree condition \cite{Riz10}. In this study, the range of
instability is independent of adiabatic index $\Gamma$ showing the
consistency of obtained results with the expansionfree condition
which requires that fluid would evolve without compressibility.
However, no such condition is imposed in this work, hence results
appeared in the form of adiabatic index.

Also, let us make the comparison of our results with Chandrasekhar's
results on the limit of dynamical instability of spherically
symmetric configuration of matter \cite{1}. In this work, dynamical
instability depends upon the numerical value of the adiabatic index.
It is found that if $\Gamma> 4/3$, the pressure in a star is strong
enough than the weight of the outer layers which make a star stable.
Whereas, for $\Gamma<4/3$, the weight increases very fast than the
pressure and star collapses resulting a dynamical instability. As
concerned to work done in this paper, results depends on the
physical quantities, like energy density, pressure, curvature terms
and mass of the cylinder. Thus, in Newtonian regime collapsing star
yield the instability limit in the the form of Eq.(\ref{74}). By
reversing the sign stability of the system would be achieved.
However, in PN regime instability range not only depends on
Eq.(\ref{PN}) but also on certain constraints in Eq.(\ref{c1}) and
(\ref{c2}).

Finally, it is mentioned here that our previous work had been done
on super gravity model while present study made on CDTT model, hence
the focus of future work is the combination of both the above
models, i.e., generalized CDTT model in any of the symmetry.

\section{Appendix}

\begin{eqnarray}\setcounter{equation}{1}\nonumber
\phi_1(r,t)&=&\frac{A^2}{\kappa}\left[\frac{f-RF}{2A^2}
+\frac{F''}{A^2B^2}-\frac{\dot{F}}{A^4}\left(\frac{\dot{A}}{A}
+\frac{\dot{B}}{B}+\frac{\dot{C}}{C}\right)
+\frac{F'}{A^2B^2}\frac{C'}{C}\right]_{,0}\\\nonumber
&+&\frac{\dot{A}}{\kappa
A}\left[\frac{f-RF}{2}+\frac{2F'}{B^2}\frac{C'}{C^2}
-\frac{\dot{F}}{A^2}\left(\frac{2\dot{A}}{A}+\frac{\dot{B}}{B}
-\frac{\dot{C}}{C}\right)\right]
\\\nonumber
&+&\frac{\dot{B}}{\kappa B}\left[\frac{\ddot{F}}{A^2}
+\frac{\dot{F}}{A^2}\left(\frac{2\dot{A}}{A}+\frac{\dot{B}}{B}\right)
+\frac{F'}{B^2}\left(-\frac{2A'}{A}
+\frac{C'}{C}\right)\right]\\\nonumber &+&\frac{\dot{C}}{\kappa
C}\left[\frac{\ddot{F}}{A^2}
-\frac{\dot{F}}{A^2}\left(\frac{2\dot{A}}{A}
+\frac{\dot{C}}{C}\right)-\frac{F'}{B^2}\frac{A'}{A}\right]+\frac{1}{\kappa
B^2}\left(\dot{F}'-\frac{A'}{A}\dot{F}\right)
\\\label{19*}&\times&\left(\frac{3A'}{A}+\frac{B'}{B}+\frac{C'}{C}\right)+
\frac{A^2}{\kappa}\left[\frac{1}{A^2B^2}\left(\dot{F}'-\frac{A'}{A}\dot{F}
-\frac{\dot{B}}{B}F'\right)\right]_{,1},\\\nonumber
\phi_2(r,t)&=&-\frac{B^2}{\kappa}\left[
\frac{1}{A^2B^2}\left(\dot{F}'-\frac{A'}{A}\dot{F}-
\frac{\dot{B}}{B}{F'}\right)\right]_{,0}\\\nonumber
&+&\left[-\frac{f-RF}{2B^2}+\frac{\ddot{F}}{A^2}+
\frac{\dot{F}}{A^2}\frac{\dot{C}}{C}
+\frac{F'}{B^2}\left(\frac{A'}{A}
+\frac{B'}{B}+\frac{C'}{C}\right)\right]_{,1}\\\nonumber
&+&\frac{A'}{\kappa A}\left[\frac{-\ddot{F}}{A^2}+\frac{F''}{B^2}
-\frac{\dot{F}}{A^2}\left(\frac{\dot{A}}{A}+\frac{\dot{B}}{B}+\frac{2\dot{C}}{C}\right)
-\frac{F'}{B^2}\left(\frac{A'}{A}+\frac{B'}{B}+\frac{C'}{C}\right)\right]\\\nonumber
&-&\frac{B'}{\kappa
B}\left[-\frac{f-RF}{2A^2}+\frac{\ddot{F}}{A^2}+\frac{\dot{F}}{A^2}
\frac{\dot{C}}{C}-\frac{F'}{B^2}\left(\frac{A'}{A}
+\frac{B'}{B}+\frac{C'}{C}\right)\right]\\\nonumber
&+&\frac{C'}{\kappa C}\left[\frac{F''}{B^2}
+\frac{\dot{F}}{A^2}\left(\frac{\dot{C}}{C}
-\frac{\dot{B}}{B}\right)+\frac{F'}{B^2}\left(\frac{2A'}{A}
+\frac{B'}{B}+\frac{C'}{C}\right)\right]\\\label{21*} &+&
 \frac{1}{\kappa A^2}\left(\dot{F}'-\frac{A'}{A}\dot{F}-
\frac{\dot{B}}{B}{F'}\right)\left(\frac{\dot{A}}{A}+\frac{2\dot{B}}{B}
+\frac{\dot{C}}{C}\right).\\\nonumber
\phi_{2s}&=&\frac{-2\mu^4R_0^{-3}}{B_0^2}\frac{A_0'}{A_0}\left[3R_0^{-1}
R_0'^2+R_0''+R_0'\left(\frac{A_0'}{A_0}+\frac{B_0'}{B_0}
+\frac{1}{r}\right)\right]
\end{eqnarray}
\begin{eqnarray}\nonumber
&-&\frac{B_0'}{B_0}\left[\mu^4R_0^{-1}+\frac{2\mu^4R_0^{-3}R_0'}{B_0^2}
\left(\frac{A_0'}{A_0}+\frac{2B_0'}{B_0}+\frac{1}{r}\right)\right]\\\nonumber
&+&\frac{2\mu^4R_0^{-3}}{B_0^2}\left(\frac{1}{r}
+\frac{B_0'}{B_0}\right)\left[3R_0^{-1}
R_0''+R_0^{-1}\left(\frac{2A_0'}{A_0}+\frac{2B_0'}{B_0}
+\frac{1}{r}\right)\right]\\\label{601}
&-&\left[\mu^4R_0^{-1}-\frac{2\mu^4R_0^{-3}R_0'}{B_0^2}
\left(\frac{A_0'}{A_0}+\frac{2B_0'}{B_0}+\frac{1}{r}\right)\right]_{,1}.\\\nonumber
\phi_{1p}&=&-\mu^4R_0^{-3}e+\frac{4\mu^4bR_0^{-3}}{B_0^4}(-3R_0'R_0^{-1}
+R_0'')-\frac{\mu^4R_0^{-4}}{B_0^2}[e''+4e'R_0^{-1}R_0'\\\nonumber
&-&4R_0^{-1}(e'R_0'-5eR_0^{-1}R_0'^2+4eR_0'')]+2\mu^4R_0^{-3}(e'R_0^{-1}-4eR_0')\\\nonumber
&\times&\left[\frac{1}{B_0^2}\left(\frac{B_0'}{B_0}+\frac{1}{r}\right)
\left(\frac{3b}{B_0}+\frac{\bar{c}}{B_0}\right)
+\frac{1}{B_0^2}\left(\frac{b+\bar{c}}{rB_0}+\frac{b'+\bar{c}'}{B_0}\right)\right]\\\nonumber
&+&A_0^2\left[\frac{1}{A_0^2B_0^2}\{2\mu^4R_0^{-4}(e'-4eR_0^{-1}R_0')
-2\mu^4R_0^{-3}\right.\\\nonumber
&\times&\left.\left.\left(eR_0^{-1}\frac{A_0'}{A_0}
+R_0'\frac{b}{B_0}\right)\right\}\right]_{,1}
-\frac{2\mu^4R_0^{-3}R_0'}{B_0^2}\frac{(b+\bar{c})}{B_0}\frac{A_0'}{A_0}\\\nonumber
&+&\frac{2\mu^4R_0^{-1}}{B_0^2}\left(e'-4eR_0^{-1}R_0'-e\frac{A_0'}{A_0}\right)
\left(\frac{3A_0'}{A_0}
+\frac{2B_0'}{B_0}+\frac{1}{r}\right)\\\nonumber
&-&\frac{a}{A_0}\left[\mu^4R_0^{-1}+\frac{2\mu^4R_0^{-3}}{B_0^2}
\left(\frac{B_0'}{B_0}+\frac{1}{r}\right)\right]\\\label{60'}
&-&\frac{2\mu^4R_0^{-3}R_0'}{B_0^2}\frac{b}{B_0}\left(\frac{2A_0'}{A_0}
+\frac{B_0'}{B_0}+\frac{1}{r}\right). \\\nonumber
\phi_3(r)&=&\frac{A_0}{\kappa
B_0}\left[\frac{1-\mu^4R_0^{-2}}{A_0B_0^2}\left\{\frac{b+\bar{c}}{r}
+b'+\bar{c}'+\frac{A_0'}{A_0} (b+\bar{c})\right\}\right.\\\nonumber
&-&\left.\frac{2A_0 \mu^4R_0^{-4}}{B_0}\left\{-e'+4eR_0^{-1}R_0'+e
\frac{A_0'}{A_0}\right\}+\frac{b}{A_0B_0^2}\right]_{,1}\\\nonumber
&-&\frac{A_0}{\kappa B_0^2}\left[\frac{1-\mu^4R_0^{-2}}{A_0
B_0}\left\{\frac{b+\bar{c}}{r}+b'+\bar{c}'
+\frac{A_0'}{A_0}(b+\bar{c})\right\}\right.\\\nonumber &-&\left.2A_0
\mu^4R_0^{-4}\left\{-e'+4eR_0^{-1}R_0'+e
\frac{A_0'}{A_0}\right\}+\frac{b}{A_0B_0}\right]\\\label{60''}
&\times&\left(\frac{2A_0'}{A_0}+\frac{B_0'}{B_0}-\frac{1}{r}\right)-\phi_{1p},
\end{eqnarray}
\begin{eqnarray}\nonumber
\phi_{2p}&=&\frac{2\mu^4R_0^{-3}}{B_0^2}\frac{\ddot{T}}{T}
\left[e'R_0^{-1}-4eR_0^{-2}R_0'
+eR_0^{-1}\frac{A_0'}{A_0}-R_0'\frac{b}{B_0}+\frac{eR_0^{-1}}{2}
\frac{B_0'}{B_0}\right]\\\nonumber
&-&\left[\frac{2\mu^4R_0^{-3}R_0'}{B_0^2}\left\{R_0'\left(\frac{a}{A_0}\right)'
+\left(\frac{b}{b_0}\right)'+\left(\frac{b+\bar{c}}{rB_0}\right)\left(2+r\frac{B_0'}{B_0}\right)
+\frac{b'}{B_0}\right.\right.\\\nonumber
&+&\left.\left.\frac{\bar{c}}{B_0}+
\frac{1}{R_0'}\left(2R_0'\frac{b}{B_0}-e'R_0^{-1}+4eR_0^{-2}R_0'\right)
\left(\frac{A_0'}{A_0}+\frac{2B_0'}{B_0}+
\frac{1}{r}\right)\right\}\right]_{,1}\\\nonumber
&+&\frac{A_0'}{A_0}\left[\frac{2\mu^4R_0^{-3}R_0'}{B_0^2}
\left(-3R_0^{-1}R_0'+\frac{R_0''}{R_0'}+\frac{A_0'}{A_0}+\frac{2B_0'}{B_0}+
\frac{1}{r}\right)\right.\\\nonumber
&\times&\left(\frac{a'}{A_0'}+\frac{a}{A_0}\right)
-\frac{2\mu^4R_0^{-4}}{B_0^2}\{e''+4e'R_0^{-1}R_0'-4R_0^{-1}
(e'R_0'\\\nonumber
&-&5eR_0^{-1}R_0'^2+4eR_0R_0'')\}+\frac{2\mu^4R_0^{-3}}{B_0^2}
\left(\frac{2R_0'}{B_0^2}+e'R_0^{-1}-4eR_0^{-2}R_0'\right)\\\nonumber&\times&
\left(\frac{A_0'}{A_0}
+\frac{2B_0'}{B_0}+\frac{1}{r}\right)-\frac{2\mu^4R_0^{-3}R_0'}{B_0^2}
\left\{\left(\frac{a}{A_0}\right)'
+\left(\frac{b}{B_0}\right)'+\left(\frac{b+\bar{c}}{rB_0}\right)\right.\\\nonumber
&\times&\left.\left.\left(2+\frac{rB_0'}{B_0}\right)+\left
(\frac{b'+\bar{c}'}{B_0}\right)\right\}\right]+\frac{B_0'}{B_0}
\left[\mu^4R_0^{-1}\left(\frac{b'}{B_0}
-\frac{b}{B_0}\right)\right.\\\nonumber
&\times&\left\{1-\frac{R_0^{-2}R_0'}{B_0^2}\left(\frac{A_0'}{A_0}
+\frac{2B_0'}{B_0}+\frac{1}{r}\right)\right\}-2\mu^4R_0^{-3}\left(\frac{A_0'}{A_0}
+\frac{B_0'}{B_0}+\frac{1}{r}\right)\\\nonumber
&\times&\left(\frac{b}{B_0^2}R_0'+e'R_0^{-1}-4eR_0^{-2}R_0'\right)-e\mu^4R_0^{-3}
-2\mu^4R_0^{-3}R_0'\\\nonumber
&\times&\left.\left\{\left(\frac{a}{A_0}\right)'
+\left(\frac{b}{B_0}\right)'+\left(\frac{b+\bar{c}}{rB_0}\right)
\right.\left.\left(2+\frac{rB_0'}{B_0}\right)+\left
(\frac{b'+\bar{c}'}{B_0}\right)\right\}\right]\\\nonumber
&+&\frac{1}{rB_0}\left[\frac{2\mu^4R_0^{-3}R_0'}{B_0^2}
\left(-3R_0^{-1}R_0'+\frac{R_0''}{R_0'}+\frac{2A_0'}{A_0}+\frac{2B_0'}{B_0}+
\frac{1}{r}\right)\right.\\\nonumber
&\times&\left((b+\bar{c})(2+r\frac{B_0'}{B_0})+r(b'+\bar{c}')\right)
-(B_0+rB_0')\left\{\frac{4\mu^4bR_0^{-3}}{B_0}(R_0''\right.\\\nonumber
&-&R_0^{-1}R_0'^2)+2\mu^4R_0^{-4}(e''+4R_0^{-1}R_0'e')-8\mu^4R_0^{-5}
(e'R_0'-5eR_0^{-1}R_0'^2\\\nonumber
&+&4eR_0'')-2\mu^4R_0^{-3}\left(\frac{2A_0'}{A_0}
+\frac{2B_0'}{B_0}+\frac{1}{r}\right)\left(e'R_0^{-1}-4eR_0^{-2}R_0'
-\frac{2R_0'b}{B_0}\right)\\\nonumber &-& 2\mu^4R_0^{-3}R_0'
\left\{\left(\frac{a}{A_0}\right)'
+\left(\frac{b}{B_0}\right)'+\left(\frac{b+\bar{c}}{rB_0}\right)
\left(2+\frac{rB_0'}{B_0}\right)\right.\\\label{61}
&+&\left.\left.\left.\left(\frac{b'+\bar{c}'}{B_0}\right)\right\}\right\}\right].
\end{eqnarray}
\begin{eqnarray}\nonumber
\phi_4(r)&=&\frac{A_0^2B_0}{2(1+b+\bar{c})}\left[e-\frac{4b}{B_0}
\left(-\frac{A_0''}{A_0}-\frac{B_0'}{rB_0}-\frac{B_0''}{B_0}
+\frac{2A_0'B_0'}{AB_0}+\frac{A_0'}{rA_0}\right.\right.\\\nonumber
&+&\left.\frac{B_0'}{rB_0}+\frac{B_0'^2}{B_0^2}\right)-
\frac{2}{B_0^2}\left\{-\frac{A_0''}{A_0}\left(\frac{a''}{A_0''}-\frac{a}{A_0}\right)
+\frac{2B_0'+rB_0''}{rB_0}\left(\frac{b+\bar{c}}{B_0}\right.\right.\\\nonumber
&-&\left.\left.\left.\frac{2b'+2\bar{c}'+b''r+\bar{c}''r}{2B_0'+rB_0''}\right)
+\frac{B_0'}{B_0}\left(\frac{a}{A_0}\right)'
+\frac{A_0'}{A_0}\left(\frac{b}{B_0}\right)'+\frac{B_0+rB_0'}{rB_0}\right.\right.\\\label{61'}
&\times&\left.\left.\left\{\left(\frac{a}{A_0}\right)'
+\left(\frac{b}{B_0}\right)'\right\}+\left(\frac{b+\bar{c}}{B_0}\right)'
\left(\frac{A_0'}{A_0}+\frac{B_0'}{B_0}\right)\right\}\right].\\\nonumber
\phi_5(r)&=&\frac{\phi_4(1-\mu^4R_0^{-2})}{\kappa}\left[\frac{2b+\bar{c}}{r}+b'+\bar{c}'
-\frac{2\mu^4R_0^{-4}(e'-4eR_0^{-1}R_0')}{1-\mu^4R_0^{-2}}\right]\\\label{75}
&-&\frac{\phi_{2p(N)}}{\kappa}. \\\nonumber
\phi_6(r)&=&\frac{\phi_4(1-\mu^4R_0^{-2})}{\kappa}\left[\frac{2b+\bar{c}}{r}+b'+\bar{c}'
+\frac{m_0(b+\bar{c})}{r^{2}}+\frac{b}{r}\left(1+\frac{m_0}{r}\right)\right.\\\nonumber
&-&\left.\frac{2\mu^4R_0^{-4}}{1-\mu^4R_0^{-2}}
\left(-e'+4eR_0^{-1}R_0'+e\frac{m_0}{r^2}\right)\right]-\frac{\phi_{2p(PN)}}{\kappa}
+\frac{\phi_4(1-\mu^4R_0^{-2})}{\kappa}\\\label{76}
&\times&\frac{2m_0}{r}\left[\frac{2b+\bar{c}}{r}+b'+\bar{c}'
-\frac{2\mu^4R_0^{-4}}{1-\mu^4R_0^{-2}} (-e'+4eR_0^{-1}R_0')\right].
\end{eqnarray}

\end{document}